# Design and Analysis of a Chaotic Micromixer with Vortices Modulation

K. Y. Tung[1] and J. T. Yang[1, 2]
[1]Department of Power Mechanical Engineering, National Tsing Hua University, Hsinchu 30013, Taiwan
[2]Institute of NanoEngineering and MicroSystems, National Tsing Hua University, Hsinchu 30013, Taiwan
d937702@oz.nthu.edu.tw, Tel: +886-3-5715131, Fax: +886-3-5724242.

*Abstract*—A novel design for vortex modulation of a passive chaotic micromixer, named a circulation-disturbance micromixer (CDM), has been achieved and analyzed experimentally and numerically. The systematic numerical analyses — topological flow characteristics and particle tracking method — have been developed, that enable visualization of detailed mixing patterns. To display the cross section of mixing region of flows in our CDM, the biotin-streptavidin binding is detected through the fluorescence resonance energy transfer (FRET) pair of fluorescent proteins — R-phycoerythrin (RPE) and cross-linked allophycocyanin (clAPC). We expect the diagnosis technique using FRET will be successfully applied to biochemical analysis in microfluidic system.
*Presentation*—Oral

## I. INTRODUCTION

Microfluidic analytic systems have been widely applied to chemical or biological assays [1] and will significantly change the ways of modern biotechnology and chemistry in the future. The miniature dimensions possessed by the microfluidic systems used in the area of lab-on-a-chip (LOC) devices, bio-micro-electro-mechanical-systems (bio-MEMS) and micro-total-analysis-systems (μTAS) [2], usually ranging from several to hundreds of micron, have turned the viscous and surface effects into the dominant factors. The macroscopic turbulent mixing is then invalid for the microscopic laminar flow. Therefore, in addition to the diffusion mechanism for laminar flow mixing, the bulk advection of the fluid must be induced to guarantee effective mixing within the microfluidic system.

Micromixers can be classified into two types: active (with external energy source for mixing) [3] and passive [4] micromixers. However, due to the simple manufacturing process and cheaper price, the passive type micromixer seems more competitive. Many researchers have proposed various chaotic micromixers by changing the structure of geometry [5] of the channel, those microchannels with patterned grooves [6],or putting on the obstacles in the channel [7].

## II. DESIGN CONCEPT AND MICRO FABRICATION

In this study, we used the tee passive micromixer consisting of two inlet ports and one outlet channel. The two different fluids is injected and mixed together on the outflow tract. The length, width and height of the mixing channel are 8100 μm, 200 μm and 70 m, respectively. The lengths of both inlet channels are 500 μm. A series of slanted grooves (width 50 μm and depth 50 μm) is patterned on the bottom wall of the mixing channel. In order to enhance the mixing of the two fluids, a zigzag barrier (width 20 μm, height 40 μm and the length of one period is 800 μm; similar to a triangular sinuous wave) was installed on the upper wall of a groove-patterned microchannel.

Our design of CDM (Fig. 1a) requires two SU-8 (SU-8 2035, MicroChem Corp.) cast moldings: one for a channel with slanted grooves on the lower plate by means of two-step photolithography, and the other for the upper plate with a zigzag barrier via one-step photolithography. We poured the degassed PDMS (polydimethylsiloxane, Sylgard 184 Silicone Elastomer, Dow Corning) mixture onto the patterned cast and peeled off the cured replicas. Next, two identical replica slabs were activated with an oxygen plasma treatment and oppositely aligned with methanol. Finally, Teflon pipes (inner diameter 460 μm, outer diameter 920 μm) were inserted into the access holes on the reservoirs to connect with disposable syringes (1 μL, with 25-gauge needles) and a syringe pump (Kd S200, Kd Scientific Inc.), as shown in Fig. 1b. To verify the performance, we have conducted systematically both numerical simulation and experimental measurement.

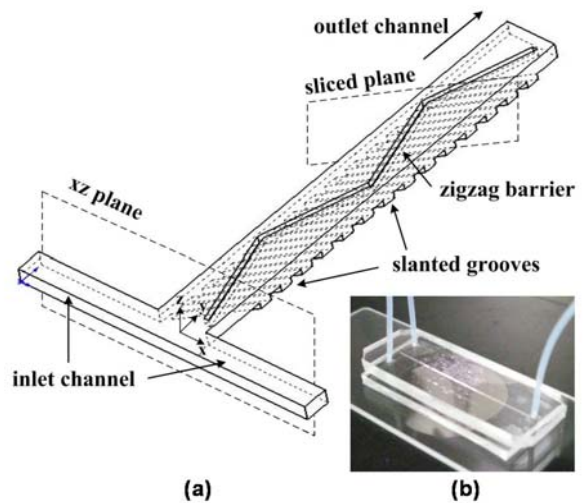

Fig. 1. (a) schematic diagram of the circulation disturbance micromixer (CDM) and (b) finished products of the CDM.

                                                                                           

## III. RESULTS AND DISCUSSION

### A. Topological flow characteristics

Fig. 2 shows the three-dimensional geometry of a mixer unit and the distributions of topological characteristics of slanted slice planes projected onto the xz plane. To achieve vortex modulation, we utilize two distinct microstructures — slanted gr to the main field; two modulated and interacting hyperbolic vortices of disparate size are thus induced. These two vortices interact across the saddle point, which is located at the middle beneath the barrier. The periodic alternations of the flow configuration and the modulation of the number, size and vorticity of vortices upgraded the extent of the flow stretching and folding, and resulted in enhanced flow mixing in the CDM.

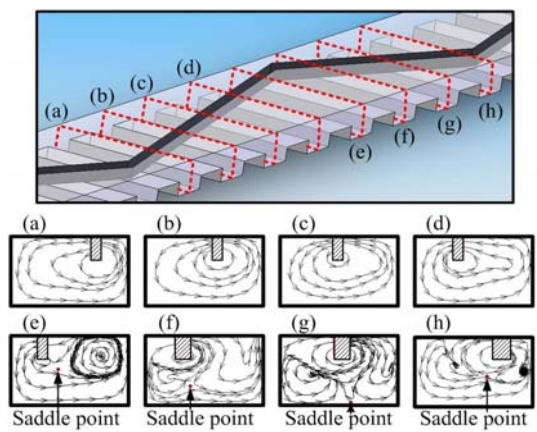

Fig. 2. Distributions of topological characteristics of slanted slice planes projected onto the xz plane (a-h). (●: saddle point).

### B. Color particle tracking

To visualize the effect on the mixing performance numerically, we placed in total 14000 passive particles of two colors (7000 per species) initially at the inlet in the computational domain of the first period such that they filled the cross section of the microchannel, as shown in Fig. 3. According to the pattern after several periods of the CDM, the flow within the mixing channel moved in a levo-rotary helical manner (viewed toward the positive-y direction). As mixing was unsatisfactory on introducing only slanted grooves (SGM) [6], we added a zigzag barrier to improve the mixing performance through an induced disturbance. From the distribution of colored particles in Fig. 3, one can easily figure out the progress of mixing from the beginning to the end of period.

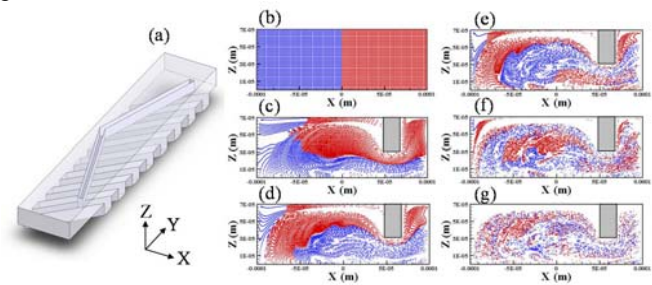

Fig. 3. Results of CFD analysis of the present mixer, (a) Geometry of a mixer unit, (b) 14000 particles uniformly seeded at the inlet. Particle distribution at the exit of (c) 1st, (d) 2nd, (e) 3rd, (f) 5th, and (g) 10th unit.

### C. The diagnosis via FRET

To diagnose the mixing efficiency in our CDM, we used the FRET method to observe a protein combination. As illustrated in Fig. 4, the FRET phenomenon involves the non-radiative transfer of radiant energy from an excited fluorophore (the donor) to another fluorophore (the acceptor) when both are located in close proximity. That the transfer is possible depends on spectral overlap and a sufficiently small distance (<10 nm) between the two fluorophores; Streptavidin R-Phycoerythrin conjugate (SA-RPE) and biotin-cross-linked allophycocyanin conjugate (BT-clAPC) served for this purpose in this FRET. Through the strong binding between biotin and streptavidin, RPE and clAPC were close enough to enable the energy transfer. For experiments with a confocal microscope, a HeNe laser excited SA-RPE at 543 nm; the fluorescent emission was selected with a 650-nm long-pass filter to detect the FRET signal.

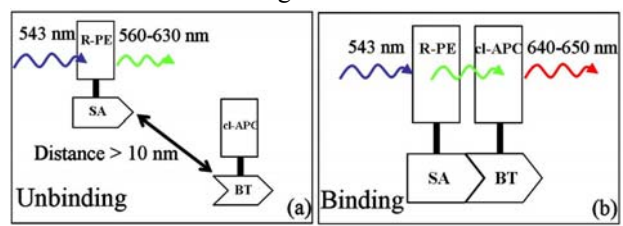

Fig. 4. Concept of the FRET method: (a) the FRET phenomenon is unactivated when the distance between samples exceeds 10 nm; (b) the FRET phenomenon is activated when the samples bind to each other. Here Streptavidin-RPE is the donor and Biotin-cl-APC is the acceptor.

Within the initial condition, FRET occurred at only the interface of two flows, caused by molecular diffusion (Fig. 5a-c). Fig. 5d, e show the results of FRET in SGM and CDM, separately. The yellow regions reveal the FRET phenomenon and are exactly the area of the reaction protein. The protein-protein binding between the donor and acceptor fluorophores increased steadily with increasing distance along the channel. In the condition of SGM, FRET occurs only from the helical motion of the fluids driven by the slanted grooves (Fig. 5d). Figure 5e demonstrates that, by effective mixing, the circulation-disturbance motion in the CDM promotes a biological reaction such as the binding of biotin and streptavidin.

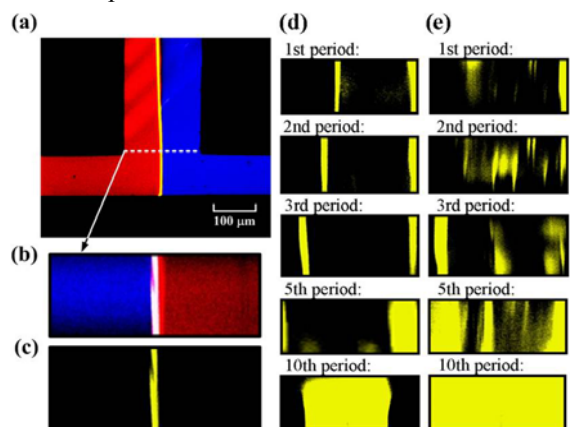

Fig. 5. Dual-track results of RPE, clAPC and FRET in SGM at a flow rate 5.0 µl min$^{-1}$ (Re ≈ 1.34): (a) xy cross-section, (b) xz cross-section. (c) Single-track result of the FRET phenomenon occurs at the interface of two flows with SA-RPE and BT-clAPC at xz cross-section, respectively. Results of FRET in (d) SGM, (e) CDM.

To quantify the above results, the FRET factor is defined as the fraction of protein reaction, RPE-clAPC, occupying the reaction channel, i.e.,

$$\text{FRET factor} = \frac{\text{Area of protein reaction}}{\text{Area of } xz \text{ cross section}} \quad (1)$$

The distribution of the FRET factor plotted to the 10th period for SGM and CDM is exhibited in Fig. 6. The agitation produced by the constructive interference of slanted grooves and zigzag barrier in CDM induces a highly significant effort for the mixing process and enlarges the mixing region. Therefore, as the contact of the two proteins increased gradually, the mutual interaction enhanced. Furthermore, the vortex modulation of a passive chaotic micromixer, CDM, raises the production rate of RPE-clAPC to 80% within six periods. Compared to the SGM, the FRET factor increases 167%.

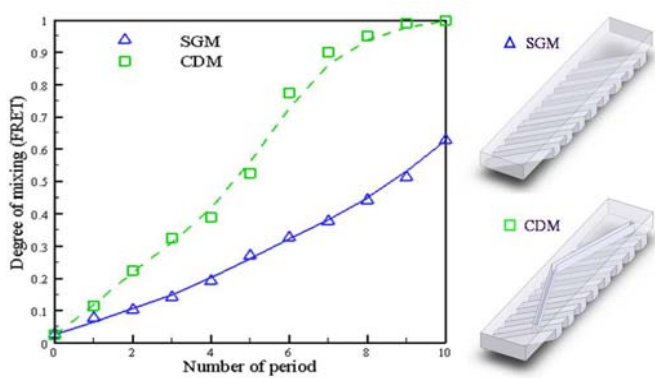

Fig. 6. The distribution of the FRET factor plotted to the 10th period for SGM and CDM.

## IV. CONCLUSIONS

In this study, systematic methods of mixing analysis were proposed to accomplish direct numerical simulation and experiment measurements of chaotic mixing in circulation-disturbance micromixer (CDM). According to the results of parallel computing, we were able to obtain the velocity flow field of micromixers and configuration of colored particles. The topological flow characteristics have been clearly illustrated that the periodic constructive interference of the vortices induced by two distinct microstructures — slanted grooves on the bottom and a zigzag barrier on the top. The particle tracking method was employed to visualize the mixing performance of two different fluids.

The fluorescence resonant energy transfer (FRET) method and a micro laser induced fluorescence (μ-LIF) system with confocal microscopy were adopted to detect the cross section of reaction zone between two fluorescent proteins, R-Phycoerythrin (RPE) and cross-linked allophycocyanin (clAPC). FRET experiment via binding of biotin and streptavidin were conducted in the CDM to verify that the vortex modulation motion, utilizing slanted grooves and a zigzag barrier, effectively promotes mixing and reaction in a microfluidic system. The mixing length is shorter than the slanted groove micromixer (SGM) (mixing efficiency is over 80% within six periods). The novel micromixer has a great potential for application in μTAS and microfluidic chip.

ACKNOWLEDGMENT

National Science Council of the Republic of China supported this work under contracts NSC 96-2628-E007-120-MY3 and NSC 96-2628-E007-121-MY3.

REFERENCES

[1] D. J. Beebe, G. A. Mensing, and G. M. Walker, "Physics and applications of microfluidics in biology," *Annu. Rev. Biomed. Eng.*, vol. 4, pp. 261-286, 2002.
[2] A. Manz, N. Grabber, and H. M. Widmer, "Miniaturized total chemical analysis systems: a novel concept for chemical sensing," *Sens. Actuators B*, vol. 1, pp. 244-248, 1990.
[3] C. C. Chang and R. J. Yang, "Electrokinetic mixing in microfluidic systems," *Microfluid Nanofluid*, in press.
[4] S. Hardt, K. S. Drese, V. Hessel, and F. Schönfeld, "Passive micromixers for applications in the microreactor and μTAS fields," *Microfluid Nanofluid*, vol. 1, pp. 108-118, 2005.
[5] R. H. Liu, M. A. Stremler, K. V. Sharp, M. G. Olsen, J. G. Santiago, R. J. Adrian, H. Aref, and D. J. Beebe, "Passive mixing in a three dimensional serpentine microchannel," *J. Microelectromech. Syst.*, vol. 9, pp. 190-197, 2000.
[6] A. D. Stroock, S. K. W. Dertinger, A. Ajdari, I. Mezić, H. A. Stone, and G. M. Whitesides, "Chaotic mixer for microchannels," *Science*, vol. 295, pp. 647-651, 2002.
[7] D. S. Kim, S. W. Lee, T. H. Kwon, S. S. Lee, "A barrier embedded chaotic micromixer," *J. Micromech. Microeng.*, vol. 14, pp. 798-805, 2004.

BIOGRAPHY

Kai-Yang Tung was born in Taiwan, R.O.C., in 1978. He received the B.S. degree from Yuan-Ze University, Chung-Li, Taiwan, in 2000 and the M.S. degree from the National Cheng Kung University, Tainan, Taiwan, in 2002. He is currently working toward the Ph.D. degree in the Department of Power Mechanical Engineering from the National Tsing Hua University, Hsinchu, Taiwan. His research interests include microfluidics system design, microfabrication, μ-LIF, and μ-PIV investigation.